\documentclass[dvipdfmx,prd,showpacs,preprint,preprintnumbers,amsmath,amssymb,eqsecnum,nofootinbib]{revtex4-1}
\usepackage{graphicx,amssymb,amsmath,amsthm,amsfonts,mathrsfs,epsfig,epsf}
\usepackage{dcolumn}
\usepackage{amsfonts,mathrsfs}
\usepackage{bm}
\usepackage{subfigure}
\usepackage{color}
\usepackage{url}
\usepackage{hyperref}
\hypersetup{
setpagesize=false,
 bookmarksnumbered=true,%
 bookmarksopen=true,%
 colorlinks=true,%
 linkcolor=blue,
 citecolor=red,
}

\usepackage[top=20truemm,bottom=20truemm,left=20truemm,right=20truemm]{geometry}

\newcommand{\D}{{\rm d}}
\newcommand{\eq}[1]{\mbox{Eq.~(\ref{#1})}}
\newcommand{\fig}[1]{\mbox{Fig.~\ref{#1}}}
\newcommand{\sign}{\text{sgn}}

\begin{document}


\title{
High energy collision without fine tuning:\\
Acceleration and multiple collisions of shells in a bound system
}

\author{Takafumi Kokubu$^{1}$} 
\email{kokubu@hunnu.edu.cn, 14ra002a@rikkyo.ac.jp}
\affiliation{$^{1}$Department of Physics and Synergetic Innovation Center for Quantum Effects and Applications, Hunan Normal University, Changsha, Hunan 410081, China}

\date{\today}          
\begin{abstract}
High energy collision of massive bodies is investigated without fine tuning.
We study multiple collisions of two spherical concentric shells in a gravitationally bound system and calculate the center of mass energy between the shells. 
We solve the equation of motions for two shells without imposing any fine tuning of the initial parameters.
In this bound system, the shells collide many times and these motions are highly nontrivial due to chaotic behavior of the shells. 
Consequently, the center of mass energy for each collision varies nontrivially and even reaches almost its theoretical upper limit. We confirm that a significant proportion of the theoretical limit is automatically achieved during multiple collisions without fine tuning.
At the same time, we also study shell ejection from the system after some collisions. If the initial shell's energy is large enough, multiple collisions may cause one shell to accumulate energy so that it escapes to infinity, even though two shells are initially confined in the system. The ejection is caused by multiple collisions inducing nontrivial energy transfer between the shells. The relation between the maximum center of mass energy and the energy transfer causing the shell ejection is also discussed.
\end{abstract}


\maketitle

\tableofcontents

\section{Introduction}
High energy collisions of massive bodies may occur around strong gravitational fields in the Universe  \cite{PiranShahamKatz1975}, and thus the collisions can be a probe for dark matter physics \cite{Baushev2008, Banados+2010}.

It has been known that strong gravitational fields work as particle accelerators.
Reference \cite{PiranShahamKatz1975} has shown that the center of mass energy, the energy between the particles in the center of mass frame, arbitrarily grows if two test particles collide head on near the horizon of an extremal Kerr black hole.  Hence, the black hole can be a natural particle accelerator. 
When the parameter of either of the particles is finely adjusted, similar energetic collision occurs \cite{BanadosSilkWest2009}, describing rear-end collision of two particles in the same spacetime as in Ref. \cite{PiranShahamKatz1975}.

Also, a black hole without rotation acts as an accelerator if two charged particles (one of them is fine tuned and the other is not) collide in the vicinity of the horizon in an extremal charged black hole spacetime \cite{Zaslavskii2010}.
Although charged black holes are not realistic in the Universe, it is often the subject of study because they have similar conformal structures to rotating black holes, but analysis in charged spacetimes is much easier than rotating ones. Therefore analysis in charged spacetimes gives a wealth of suggestions for more realistic spacetimes with rotation.

One of criticisms on the high energy collision due to strong gravity is that the backreaction by self-gravity of the colliding object is not taken into account, thereby lacking sufficient accuracy for the motion of the object near strong gravitational regions. However, solving the Einstein equations including self-gravity is in general hard task.
To bypass the difficulty, the shell model implemented by Israel is often adopted \cite{Israel1966}.
A thin-matter layer confined on a singular hypersurface is called a shell (or brane). In the Israel's formalism,  the self-gravity of the shell is fully taken into account, i.e., the shell solves the Einstein equations.
As a result, it was revealed that the center of mass energy has an upper bound when the back-reaction is taken into account by considering collision of two charged shells (one particle is fine tuned) in the extremal Reissner-Nordstr\"om spacetime \cite{KimuraNakaoTagoshi2010}.

An another criticism on high energy collisions is the difficulty of ``fine tuning''.
Two kinds of fine tunings on the initial conditions may be required for high energy collisions to occur: 
(i) one of the colliding bodies being ``critical,'' meaning that the body is tuned to satisfy an algebraic relation between physical quantities proper to the body (e.g., conserved energy, angular momentum or charge, in the case of test particles) and inertial quantities of the background spacetime (e.g., the rotation of the spacetime), and 
(ii) tuning of the initial position and the initial velocity of each body so that the critical body collides with the other noncritical body just above the horizon. 

Meanwhile, we note that large amount of the center of mass energy is realizable without black holes.
An overextremal Reissner-Nordstr\"om spacetime containing no horizon, but a naked singularity, works as a high energy  accelerator \cite{Patil+2011}.
In an overextremal Reissner-Nordstr\"om spacetime that is very close to an extremal spacetime, the center of mass energy becomes arbitrarily large when the two neutral particles or the two neutral shells collide at a particular radius. \footnote{More precisely, the ratio of the center of mass energy to the proper mass becomes arbitrarily large.} 
An interesting feature in this system is that we do not need a fine tuning of the initial parameters. The fine tuning listed in the above as i is not required for obtaining the high energy.
Although this sounds fascinating, the fine tuning of type ii is still required. The particles (or shells) must be dropped from a distant region in a timely manner in order that the particles (shells) collide at the very particular collision point, causing high center of mass energy.

Nonetheless, it is very fascinating if high energy collision occurs without artificial tuning. In this respect, motion of two objects in bound systems is perhaps useful for realizing multiple collisions without fine tuning. Because the shells are confined and they collide many times, we may expect that some collisions become highly energetic if they collide at the very particular collision point.

In our real Universe, massive objects in self gravitating systems (e.g., constituent particles in a star, or stars in clusters) collide with each other many times due to self-gravity and velocity dispersion. They contract due to self-gravity and expand due to velocity dispersion, and then contract again. 
This routine may cause high energy collision without fine tuning if they collide at a critical radius responsible for high center of mass energy.




Behavior of multiple shells has been investigated not only in gravitational physics \cite{Israel1966} - \cite{NakaoYooHarada2018} but also in the context of astrophysics. By tracking motions of $N$ shells, time evolution and the end state of spherically symmetric gas cloud were studied by Henon  \cite{Henon1967}. 
The mass distribution of spherical cluster was explained by solving $N$ shell dynamics in Newtonian mechanics \cite{Yangurazova+1984}.
As discussed in these references, some of shells gain a large amount of energy by exchanging energy between shells due to collisions, thereby causing them to escape from the system. This is interpreted as mass ejection from the system, which is confirmed in Newtonian \cite{MillerYoungkins1997, Barkov+2001MN} and in general relativity \cite{Barkov+2002JTEP}.
On the other hand, when mass ejection does not occur and the shells are confined in the system for long time, gravitating shells in general exhibit chaotic behavior even in Newtonian gravity \cite{MillerYoungkins1997, Barkov+2001MN}.
Motion of two concentric shells with identical mass moving between an inner and outer reflecting spherical box is chaotic in Newtonian mechanics \cite{MillerYoungkins1997}.
The similar situation but with different masses around a central body in Newtonian gravity is also chaotic \cite{Barkov+2005}.

Under these circumstances, in this paper we consider motions of two electrically neutral dust shells that are initially gravitationally bound in the overextremal Reissner-Nordstr\"om  spacetime containing the naked singularity. 
In our setup, dust shells are initially confined in a finite range of radius due to self-gravity of the shells and repulsive force by the naked singularity at the center. As a result, this simple setup naturally makes a gravitationally bound system. 
Since multiple collisions take place when the shells are confined, we consider if large amount of the center of mass energy can be achieved during the collisions.

We assume that the shells are ``transparent'', i.e., they interact only gravitationally so that they pass through during collision.
Note that a candidate of transparent matter is dark matter, interacting only gravitationally. 
This simple assumption naturally induces multiple collisions in the confining geometry.
The treatment of transparent shells has been already formulated in Refs. \cite{Nunez+1993, NakaoIdaSugiura1999, IdaNakao1999, Barkov+2002JTEP}.
This formalism was applied to a two-shell system describing the critical behavior of black hole formation in asymptotically flat \cite{CardosoRocha2016} and also in asymptotically anti-de Sitter spacetime \cite{BritoCardosoRocha2016}. Another application is charged two shells moving around a charged black hole, describing perpetual oscillating motions \cite{MazharimousaviHalilsoy2018}. 
Adopting the formalism, we numerically solve equations of motions of two dust shells in a bound system under generic initial conditions, thereby forcing induction of multiple collisions. We calculate the center of mass energy between the shell at each collision to evaluate how large the energy is.

Here, we comment on spacetimes with naked singularities. 
Spacetimes containing naked singularities are unpleasant, because theory of general relativity breaks down.
The weak cosmic censorship conjecture \cite{Penrose1969} was thus introduced to forbid the naked singularity from generating by physically reasonable initial conditions.
However, counterexamples to the conjecture have been already reported: inhomogeneous spherical dust collapse leads to the forming of a locally naked singularity \cite{JoshiDwivedi1993}. This is also true even though collapsing matter possesses small pressure \cite{OriPiran1987}.
Moreover, a spacetime with a naked singular region would be effectively realizable if one considers a quantum gravity that may resolve the infinite curvature of the naked singularity, indicating that classical naked singular spacetimes are indeed appropriate solutions except the central singularity \cite{Johnson+1999}. 
Considering above, physical reality of naked singular spacetimes is still an open issue.
Therefore, in this paper we do not pursue the question for reality of naked singularity. We introduce a naked singularity just as an origin of repulsive force causing bounce of contracting bodies, rather than physical entity. Once we accept the naked singularity, this system, two shells in the overextremal Reissner--Nordstr\"om spacetime, provide a very simple confining geometry.

This paper is organized as follows. In Sec. \ref{sec:setup}, we construct dust shells in the overextremal Reissner--Nordstr\"om spacetime and review the collisions of transparent shells.
In Sec. \ref{sec:results}, we numerically solve two shell problems under generic initial conditions, confirming that high energy collisions indeed occur.  
A summary and conclusion are devoted to  Sec. \ref{sec:conclusion}.

We take the gravitational constant $G$ and the speed of light $c$ are unity.

\section{Setup}
\label{sec:setup}
In this section, we review collisions of two, electrically neutral dust shells in the overextremal Reissner--Nordstr\"om spacetime.

The Reissner--Nordstr\"om spacetime is the unique, static and spherically symmetric solution to the Einstein-Maxwell equations with the potential 1-form, $A_\mu \D x^\mu=-(Q/r)\D t$, whose metric is written by
\begin{align}
\D s^2=-f(r)\D t^2+f(r)^{-1}\D r^2+r^2( \D \theta^2+\sin^2\theta\D \phi^2)
\label{ds}
\end{align} 
with
\begin{align}
 f(r)=&1-2M/r+Q^2/r^2.
\end{align}
The parameter $M$ is the gravitational mass (the Misner-Sharp energy) and $Q$ is the charge of the spacetime. 
This solution has a curvature singularity at $r=0$.
Since we focus on an overextremal solution, $M<Q$.
Combining the inequality and the assumption that  $M$ is positive, we consider the spacetime with
\begin{align}
0<M<Q \label{extremal-condition}
\end{align}
throughout this paper.
The condition (\ref{extremal-condition}) guarantees that $f(r)>0$ which means the event horizon is absent and the central curvature singularity is naked, i.e., an asymptotic observer at infinity would see the singularity.

\subsection{Single dust shell}
Before constructing the two-shell system, we first introduce a single shell residing on a timelike hypersurface $\Sigma$, which partitions the spacetime into the inner ($-$) and the outer ($+$) regions.
The following shell-construction is based on Ref. \cite{Poisson2009}.
The Einstein-Maxwell equations for the shell are calculated by Israel's junction conditions.
On the hypersurface the line element is given by
$\D s_{\Sigma}^2=h_{ab}\D y^a \D y^b=-\D \tau^2+R(\tau)^2 (\D \theta^2+\sin^2\theta\D \phi^2)$ with the shell's radius $R(\tau)$, where $\tau$ is the proper time of the shell.  
$\{y^a\}$ is the intrinsic coordinates on $\Sigma$ and  is chosen as $y^a=(\tau, \theta, \phi)$.
The unit normal $n^\alpha$ to $\Sigma$ and the basis vectors $e^\alpha_{a}:=\partial x^\alpha/\partial y^a$ tangent to $\Sigma$ are written by 
$n_{\alpha \pm}\D x^\alpha = -\dot{R}\D t+ \dot{t}_\pm \D r$, $u_\pm^\alpha \partial_{\alpha}:= e_{\tau \pm}^\alpha \partial_{\alpha}=  \dot{t}_\pm \partial_{t}+\dot{R} \partial_{r}$, $e^\alpha_{\theta}\partial_\alpha =\partial_{\theta}$, and $e^\alpha_{\phi}\partial_\alpha =\partial_{\phi}$.
$u^\alpha$ and $n^\alpha$ are normalized and orthogonal each other: $u^{\alpha}u_{\alpha}=-1$, $n_\alpha n^\alpha=1$ and $u^\alpha n_\alpha=0$.
We have defined $\dot{} := \partial/\partial \tau$. 
The subscript with the plus sign (the minus sign) denotes quantities in the outer (inner) region.
Junction conditions are written as $[h_{ab}]=0$ and  
\begin{align}
8\pi S_{ab}=- [K_{ab}]+[K] h_{ab}, \label{hypersurface-eom}
\end{align}
where $K_{ab}:=n_{\alpha;\beta}e^\alpha_a e^\beta_b$ is the extrinsic curvature and $S_{ab}$ is the stress-energy tensor of the shell. We defined a gap of tensorial quantities at the hypersurface, $[X]:=(X^+-X^-)|_{\Sigma}$.
The constraints equations are given by
\begin{align}
&S^{~b}_{a~|b}=- [T_{\alpha\beta}e^\alpha_{a}n^\beta], \label{hamiltonian-cons} \\
&\bar K^{ab}S_{ab}=  [T_{\alpha\beta}n^\alpha n^\beta], \label{momentum-cons}
\end{align}
where $\bar K^{ab}:=(K^{ab}_++K^{ab}_-)|_\Sigma/2$ and $X_{|a}$ is the covariant derivative with respect to the induced metric $h_{ab}$.
The bulk stress-energy tensor $T^\alpha_{~\beta}$ is given by $T^\alpha_{~\beta}=Q^2/(8\pi r^4) {\rm diag} (-1,-1,1,1)$.
From one of the junction conditions, $[h_{ab}]=0$, $\dot t_\pm$ has a relation as
\begin{align}
\dot t_\pm:=\frac{\beta_\pm}{f_\pm(R)}, \quad \beta_\pm:=\sqrt{f_\pm (R)+\dot{R}^2}.
\label{t-dot-root}
\end{align}

The nonzero components of the extrinsic curvature are 
$K_{\tau \pm}^{\tau}=\dot \beta_\pm/\dot{R}$ and $K_{\theta \pm}^{\theta}=K_{\phi \pm}^{\phi}=\beta_\pm/R$.
We take $S^i_j$ as a perfect fluid form, $S^i_j={\rm diag}(-\rho,p,p)$ with the surface pressure $p$ and the surface energy density $\rho$.
Then, the junction condition \eq{hypersurface-eom} reduces to
\begin{align}
-4\pi \rho&=(\beta_+- \beta_-)/R, \label{JUNCTION1} \\
8\pi p&=(\dot \beta_+-\dot \beta_-)/\dot{R}+(\beta_+- \beta_-)/R. \label{JUNCTION2}
\end{align}

\eq{hamiltonian-cons} is explicitly  written by
\begin{align}
R\dot{\rho}=-2\dot{R}(p+\rho). \label{var-energy-cons}
\end{align}
From \eq{var-energy-cons}, $\rho$ is solved as $\rho=\rho(R)$ when the equation of state is given. 
As mentioned, since we suppose the shell is made of dust fluid, $p=0$. Then, \eq{var-energy-cons} is integrated to give
\begin{align}
m:=4\pi R^2 \rho.
\end{align}
$m$ is a constant and denotes the proper mass of the shell. We assume $m>0$ throughout this paper.

By squaring \eq{JUNCTION1} twice to eliminate the square-root term, we arrive 
the energy equation for an electrically neutral dust shell,
\begin{align}
&\dot R^2+V(R)=0, \nonumber \\
&V(R)=1 -E^2-\frac{M_++M_-}{R} +\frac{Q^2}{R^2} -\left(\frac{m}{2R}\right)^2,
\label{dust-potential}
\end{align}
where
\begin{align}
E:=(M_+-M_-)/m.
\end{align}
$M_+-M_-$ denotes the shell's Misner-Sharp energy and $E$ is the shell's specific energy.
\eq{dust-potential} describes the dynamics of the shell. The fifth term, $-(m/2R)^2$, in \eq{dust-potential} corresponds to the self-gravity of the shell. The shell can move within a range where $V(R)\leq 0$. 
$E$ has a positive lower bound $E_c$, where the local minimum of the potential \eq{dust-potential} touches the $R$ axis, meaning that $V=\D V/\D R=0$, which is solved in terms of the specific energy $E$ as
\begin{align}
E_c=\sqrt{1-\frac{(M_++M_-)^2}{4Q^2-m^2}}. \label{Ec}
\end{align}
Note that for large radius, $V(R\rightarrow \infty)=1-E^2$. Hence, the shell is bounded if $E_c<E<1$ (bound motion), the shell reaches infinity with vanishing velocity if $E=1$ (marginally bound motion), while the shell reaches infinity with nonvanishing velocity if $E>1$ (unbound motion).

For small radius, $R^{-2}$ terms in \eq{dust-potential} become dominant, and as a result a contracting shell can bounce back if the shell's proper mass is smaller than twice the charge,
\begin{align}
m<2Q. \label{bounce-RN}
\end{align}
On the other hand, for $m\geq 2Q$, a contracting shell collapses to a black hole because there is no inner potential barrier.
Throughout this paper, we assume the shell with $E_c<E<1$ and $m<2Q$ so that the shell is gravitationally bound.

For later convenience, we rewrite \eq{t-dot-root} in a simpler form without the square root.
By multiplying the both sides of \eq{JUNCTION1} by $(\beta_++\beta_-)$, we obtain
\begin{align}
\beta_+\beta_-=-2\left(\frac{m}{2R}\right)^2+\frac{f_++f_-}{2}-V.
\label{betabeta}
\end{align}
By using \eq{betabeta} to eliminate the $\beta_+\beta_-$ term in \eq{JUNCTION2}, we obtain an algebraic relation between $\beta_+$ and $\beta_-$. With the help of the algebraic relation and \eq{JUNCTION1},
 we solve $\beta_\pm$ without the square root as
\begin{align}
\beta_\pm=\mp\left(\frac{m}{2R}\right)-\frac{f_+-f_-}{4}\left(\frac{2R}{m}\right).
\end{align}
Thus we arrive the expression without the square root,
\begin{align}
f_\pm \dot t_\pm=E \mp \frac{m}{2R}. \label{time-dilation}
\end{align}

\subsection{Two shells and collisions}
\begin{figure}[htbp]
  \begin{center}
    \begin{tabular}{c}

      \begin{minipage}{0.5\hsize}
        \begin{center}
          \includegraphics[scale=0.9]{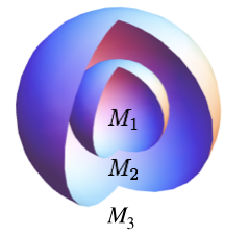}
          (a) 
        \end{center}
      \end{minipage}
      
      \begin{minipage}{0.5\hsize}
        \begin{center}
          \includegraphics[scale=1]{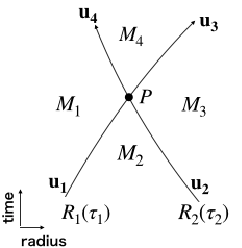}
         (b) 
        \end{center}
      \end{minipage}
 
\\
    \end{tabular}
    \caption{(a) Two-shell system. (b) Schematic picture of shell collision. The inner shell with the radius $R_1$ and the outer with $R_2$ collide at point $P$.}    
\label{fig-collision}
  \end{center}
\end{figure}

We have introduced the single shell.
Now, we introduce ``two'' shells and these collisions.
The two-shell system consists of two concentric shells. The shells divide the spacetime into four regions, say, the region $I$ ($I=1,2,3,4$). Each region possesses different gravitational mass, $M_I$.
See \fig{fig-collision} for the configuration of the two-shell system.
We use \eq{dust-potential} to track the motion of the inner shell with the radius $R_1$ and the outer one with $R_2$.
The equation of the inner shell is obtained in \eq{dust-potential} by setting $M_+=M_2$ and $M_-=M_1$. Similarly, motion of the outer shell can be obtained by setting $M_+=M_3$ and $M_-=M_2$.  
For simplicity we assume the proper mass for both shells is the same, i.e., $m_1=m_2=m$. 

Generally, when shells collide with each other, so-called the shell-crossing singularity forms there, and subsequent analysis becomes impossible unless appropriate boundary conditions are imposed at the collision event. One of such conditions is a transparent condition \cite{Nunez+1993, NakaoIdaSugiura1999} which consists of two conditions:
(1) The four-velocity of each shell is continuous at the collision.
(2) The proper mass $m$ of each shell is invariant during the collision.
In other words, each shell just goes through the other. 
Although the four-velocity is continuous, acceleration of shells is discontinuous because the gravity that the shells feel varies discontinuously before and after the collision. That is, the gravitational mass (the Misner-Sharp energy) between the shells varies discontinuously at the collision, namely, $M_2 \rightarrow M_4$ (see \fig{fig-collision}). 
The explicit form of $M_4$ after the collision is given in Refs. \cite{NakaoIdaSugiura1999, IdaNakao1999}. 
In our setup, by letting $P$ be the point at the collision event, $M_4$ is written as
\begin{align}
M_4|_P=&M_3-M_2+M_1+\frac{1}{Rf_2}\left(M_2-M_1-\frac{m^2}{2R}\right)\left(M_3-M_2+\frac{m^2}{2R}\right) \nonumber \\
& -\frac{\sign(\dot R_1)\sign(\dot R_2)}{Rf_2}\sqrt{\left(M_2-M_1-\frac{m^2}{2R}\right)^2-m^2f_2}\sqrt{\left(M_3-M_2+\frac{m^2}{2R}\right)^2-m^2f_2}.
\label{M2after}
\end{align}
The right-hand side of \eq{M2after} is evaluated at the collision event $P$.
We defined $\dot R_{1,2}:=\D R_{1,2}/\D \tau_{1,2}$.
Equivalently, the metric component $f_2$ in region 2 varies as $f_2 \rightarrow f_4$ at the collision, where
\begin{align}
f_4=1-2M_4/R+Q^2/R^2
\end{align}
is the metric component in region 4. 
After the collision, \eq{dust-potential} is still applied to track dynamics of each shell merely by replacing $M_2$ with $M_4$. 

Since the proper time for each shell is in general different, we must use a time coordinate common to the two shells to follow long time evolution and multiple collisions. A convenient choice for the common time coordinate is $t_2$ which is the time measured between the shells.
By multiplying \eq{dust-potential} by \eq{t-dot-root}, we have
\begin{align}
\left(\frac{\D R_i}{\D t_2}\right)^2+\frac{f_2(R_i)^2V(R_i)}{f_2(R_i)-V(R_i)}=0 \qquad (i=1, 2).
\label{commontime-potential}
\end{align}

\subsection{Center of mass energy}
\label{sec:CMenergy}
In order to measure how the collision of shells is energetic, we introduce the the center of mass energy.
The the center of mass energy $E_{cm}$, the energy between the two shells for their center of mass frame, has been given by \cite{KimuraNakaoTagoshi2010}
\begin{align}
E_{cm}^2=2m^2(1+\gamma), \quad
\gamma:=-g_{\alpha \beta}u_1^\alpha u_2^\beta. \label{Ecm}
\end{align}
As assumed, both shells have the same proper mass.
$\gamma ~(\geq 1)$ is the Lorentz factor of the relative velocity between the shells and can be calculated with the metric in  region 2 as $\gamma=-f_2\dot t_1 \dot t_2+f_2^{-1}\dot R_1 \dot R_2$.
From \eq{Ecm}, $E_{cm}/m$ is large when the relative velocity between the shells is large. 
There is no upper limit on the maximum value of $E_{cm}/m$ because the shells can collide at a relativistic velocity ($\gamma\gg1$) due to strong gravitational fields.
On the other hand, there is the lower limit for the center of mass energy, $\min \{E_{cm}\}=2m$, when $\gamma=1$. 
Note that $\gamma\simeq 1$ when the Newtonian approximation is valid. Thus, $E_{cm}/2m$ measures the increase in the center of mass energy due to the relativistic effect.
It is important to explain for more details on the meaning of the ratio $E_{cm}/2m$. Assuming that the shell consists of many particles, the ratio is also interpreted as the center of mass energy per unit mass of the constituent particles. If all the constituent particles have an identical mass $\tilde m$, the number of particles contained in one shell is given by $n=m/\tilde m$. Then, the center of mass energy of the constituent particle is given by $e=E_{cm}/2n$. Thus, the center of mass energy per unit mass of the constituent particle is given by $e/\tilde m=E_{cm}/2m$.
Therefore, a large $E_{cm}/2m$ means a large collisional energy between the constituent particles. 

With the help of Eqs. (\ref{dust-potential}) and (\ref{time-dilation}), the explicit form of \eq{Ecm} is written by
\begin{align}
\frac{E_{cm}^2}{2m^2}=&
1+\frac{1}{f_2}\left(E_1-\frac{m}{2R}\right)\left(E_2+\frac{m}{2R}\right) \nonumber \\
& -\frac{\sign(\dot R_1)\sign(\dot R_2)}{f_2}\sqrt{\left(E_1-\frac{m}{2R}\right)^2-f_2}\sqrt{\left(E_2+\frac{m}{2R}\right)^2-f_2},
\label{Ecm-explicit}
\end{align}
where $\sign(x)$ is the sign function.
It seems that \eq{Ecm-explicit} can be large if $f_2$ is small for given values of $E_i, m, Q$ and $M_2$. Since the spacetime is overextremal in our setup, $f_2$ is positive and never becomes zero. The minimum value of $f_2$ is 
\begin{align}
f_{2min}(R_{min})=1-M_2^2/Q^2 \quad {\rm at} \quad R_{min}=Q^2/M_2.
\label{f2}
\end{align}
When $Q\rightarrow M_2$, $f_2$ approaches $0$ from above, $f_2\rightarrow 0$.
When $f_2$ is close to zero, if the collision is rear end $[\sign(\dot R_1)\sign(\dot R_2)=+1]$, a large center of mass energy cannot be achieved because $f_2^{-1}$ terms in the second and the third terms in \eq{Ecm-explicit} cancel out.
On the other hand,  the largest $E_{cm}$ is possible if and only if the shells collide head on $[\sign(\dot R_1)\sign(\dot R_2)=-1]$ exactly at $R=R_{min}$. 
Having said that, the largest collision is unlikely unless we impose a fine tuned initial position and velocity  of the shells so that they collide miraculously at $R=R_{min}$.

For later use, we analytically evaluate the largest value of the center of mass energy, that is achieved when the shells are ``fine tuned.''
The center of mass energy becomes large when a head-on collision takes place at $R=R_{min}$. 
Recalling that $M_2$ is essentially a variable (it varies at each collision), we evaluate the maximum value of $E_{cm}|_{R_{min}}$ by regarding it as a function of $M_2$. 
However, a straightforward calculation shows that the explicit function of $E_{cm}|_{R_{min}}$ obtained by substituting $R=R_{min}$ into \eq{Ecm-explicit} is not analytically solvable in terms of $M_2$. 
In order to treat the function analytically, we notice the quantity $E_{cm}|_{R_{min}}$ with $\sign(\dot R_1)\sign(\dot R_2)=-1$ satisfies the following inequality.
\begin{align}
\frac{E_{cm}^2}{2m^2}=&
1+\frac{1}{f_2}\left(E_1-\frac{m}{2R}\right)\left(E_2+\frac{m}{2R}\right)
 +\frac{1}{f_2}\sqrt{\left(E_1-\frac{m}{2R}\right)^2-f_2}\sqrt{\left(E_2+\frac{m}{2R}\right)^2-f_2}  \nonumber \\
<&1+\frac{1}{f_2}\left(E_1-\frac{m}{2R}\right)\left(E_2+\frac{m}{2R}\right) 
 +\frac{1}{f_2}\left(E_1-\frac{m}{2R}\right)\left(E_2+\frac{m}{2R}\right)  \nonumber \\
=&1+\frac{2}{f_2}\left(E_1-\frac{m}{2R}\right)\left(E_2+\frac{m}{2R}\right)=: F(R). 
\label{Ecm-theoretical-max}
\end{align}
We used $f_2>0$ and the positivity of $\dot t$ by \eq{time-dilation} in the first inequality. 
When the collision happens at $R_{min}$, not only $E_{cm}$ but also $F(R_{min})$ take maximum values for given values of $E_i, m, Q$, and $M_2$.
Thus, $F(R_{min})$ can be used to measure the upper bound of $E_{cm}$ if $f_{2min}$ is sufficiently close to 0. Indeed, we will consider such a situation ($f_{2min}\simeq0$) in the next section.
Focusing on the form of $F(R_{min})$, it can be regarded as a function of $M_2$.
Then, this function has one local maximum at $M_2=M_{2ex}$ in the range $M_1<M_2<M_3$, 
where $M_{2ex}$ satisfies $\partial F(R_{min})/\partial M_2|_{M_{2ex}}=0$ and is simply given by
\begin{align}
M_{2ex}=\frac{M_1M_3+A^2Q^2}{A(M_1+M_3)}\left\{ 1-\sqrt{1-\left(\frac{AQ(M_1+M_3)}{M_1M_3+A^2Q^2} \right)^2}\right\}, \quad A:=1-m^2/(2Q^2).
\end{align}
$A$ is positive if $m$ is sufficiently smaller than $Q$, and this is indeed the case that we consider in the next section.
Now, we define an analytic upper bound of the center of mass energy, $E_{cm. b}/2m$, by substituting  $M_2=M_{2ex}$ into $F(R_{min})$,
\begin{align}
\frac{E_{cm. b}}{2m}:=\left.\sqrt{\frac{F(R_{min})}{2}}\right|_{M_{2ex}}
=\sqrt{\frac{1}{2}\left\{1+\frac{A^2Q^2}{m^2}\left(2A-\frac{M_1+M_3}{M_{2ex}}\right)\right\}}.
\label{thmax-Ecm}
\end{align}
\eq{thmax-Ecm} is the theoretical maximum value of $E_{cm}/2m$, that is achieved when the shells are fine tuned.

\subsection{Energy transfer}
We review the energy transfer between shells \cite{NakaoIdaSugiura1999}.
The shell transfers its energy to the other shell when they collide.
Let $\Delta E$ be the value of the energy transfer. 
By the use of \eq{M2after} and recalling that the specific energy of each shell after the collision is given by $E_3=(M_3-M_4)/m$ for the outer shell and $E_4=(M_4-M_1)/m$ for the inner shell, we write the energy transfer between shells at the collision by
\begin{align}
E_3=E_1-\Delta E \quad {\rm and} \quad
E_4=E_2+\Delta E, 
 \label{energy-transfer}
\end{align}
where 
\begin{align}
\Delta E=\gamma m/R_c. \label{delta-E}
\end{align}
$R_c$ is the radius of the collision point $P$ [see \fig{fig-collision} (b)].
\eq{delta-E} denotes that $\Delta E>0$ and the energy transfer is large when the Lorentz factor is large.
The minimum energy transfer $\Delta E\simeq m/R_c$ is realized in the Newtonian regime $\gamma\simeq 1$. In other words, relativistic motions necessarily increase the energy transfer.

From \eq{energy-transfer} and \eq{delta-E}, a universal feature for the collision is immediately revealed: At the collision, the inner shell (with $E_1$) always releases its energy $\Delta E$, whereas the outer shell (with $E_2$) always gains $\Delta E$.
From \eq{energy-transfer}, we find the energy conservation,
\begin{align}
E_1+E_2=E_3+E_4.
 \label{energy-cons}
\end{align}

\section{Results}
\label{sec:results}
In this section, we numerically solve the equations of motion for two shells without imposing fine tuned initial parameters on the shells and show that high energy collision indeed occurs after some collisions under generic initial conditions.

\subsection{Initial conditions}
\label{sec:initialcondition}
Since there are lots of parameters (gravitational masses in the regions 1, 2, and 3; the initial gravitational and specific energies; the radius; and the initial direction for each shell), we investigate situations with the following assumptions:
\begin{itemize}
\item[(i)] The overextremal spacetime we consider satisfies
\begin{align}
Q=(1+\epsilon)M_1, \quad 0<\epsilon \ll1.
\label{almost-extremal}
\end{align}
The spacetime in region 1 becomes close to the extremal Reissner--Nordstr\"om solution as  $\epsilon \rightarrow 0$. 
\item[(ii)] Let $\mu_i$ be the shell's gravitational mass. We take $M_2=M_1+\mu$ and $M_3=M_1+2\mu$ so that the shells have the same gravitational mass, $\mu=\mu_i$.
\item[(iii)]  As the initial parameters, the gravitational masses in  regions 1, 2, and 3 and the charge $Q$ satisfy
\begin{align}
M_1<M_2<M_3<Q
\label{monotonic-mass}
\end{align}
in order to ensure that  regions 1, 2. and 3 are overextremal ($M_{1,2,3}<Q$), implying that a black hole never forms by collisions. 
As a consequence of the combination of \eq{almost-extremal} and the inequality (\ref{monotonic-mass}), $\mu$ must satisfy $\mu<\epsilon M_1/2$, denoting that the shell's gravitational mass is at most $\mathcal O(\epsilon)$. 
In other words, we are considering a situation in which the shell's gravitational mass is too small to collapse into a black hole. 
Then, $\mu$ is naturally parametrized by introducing a parameter $\delta$ as
\begin{align}
\mu=\epsilon M_1 \delta/2,  \quad (0<\delta<1).
\end{align}
We fix the shell's gravitational mass $\mu$ via $\delta=0.9$ to reduce a great number of freedoms in the initial parameters.
\item[(iv)] Each shell has the same specific energy at the initial time, $E_0:=E(t=0)=E_i(0)$.
\item[(v)]  The two shells start from the same initial radius, $R_0:=R(t=0)=R_i(0)$.
\item[(vi)] The shell with the radius $R_2$ initially moves outwardly, $\sign(\dot R_2(0))=+1$.
\end{itemize}

Summarizing above, the initial parameters that we take are $\epsilon, E_0, R_0$, and $\sigma_1:=\sign(\dot R_1(0))$. 
The value of the proper mass $m$ is identified via $E_0=\mu/m$ when $E_0$ and $\mu$ are specified.
From now on, without losing generality, we take $M_1=1$.

Let us recall that the center of mass energy is given by \eq{Ecm-explicit}. With the assumption of \eq{almost-extremal}, we can estimate the dependence of $\epsilon$ on $E_{cm}$ as 
\begin{align}
E_{cm}/m \propto \epsilon^{-1/2}.
\label{Ecm-order}
\end{align}

\subsection{High energy collision without fine tuning}
Here, we solve \eq{commontime-potential} to follow the time evolution of the two shells.
The equations are so simple that they can be numerically integrated merely by using NDSolve, a built-in command in {\it Mathematica} (with 40 digits of Working precisions in our calculation).

\begin{figure*}[htbp]
  \begin{center}
    \begin{tabular}{c}

      \begin{minipage}{0.5\hsize}
        \begin{center}
          \includegraphics[scale=0.7]{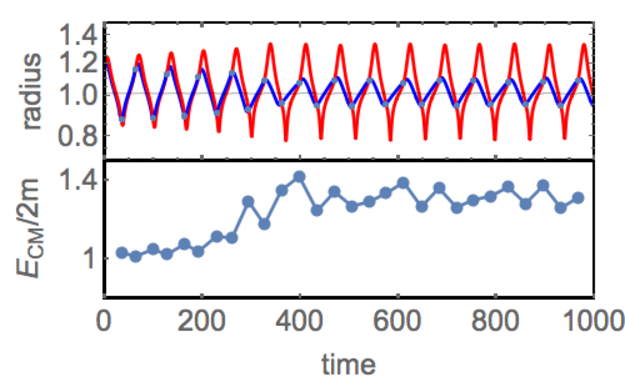}
          (a) 
        \end{center}
      \end{minipage}
      
      \begin{minipage}{0.5\hsize}
        \begin{center}
          \includegraphics[scale=0.7]{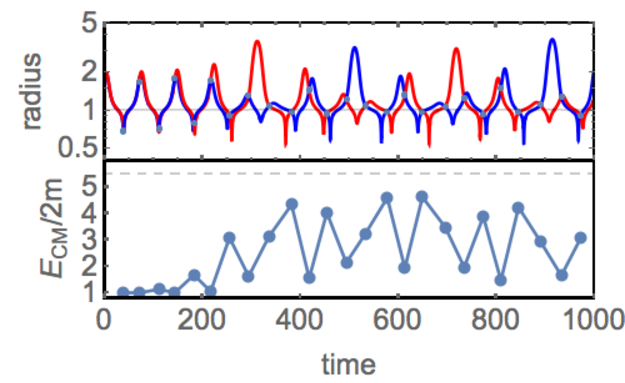}
         (b) 
        \end{center}
      \end{minipage}
 
\\
    \end{tabular}
    \caption{(a) Upper panel: The evolution of the shells in the bound system with $\epsilon=10^{-2}, E_0=0.2, R_0=1.2$, and $\sigma_1=1$. Two shells are depicted as the red and the blue curves. The black line denotes $R_{min}$ at which the maximum collision (the fine tuned collision) takes place. Twenty-three collisions occur up to $t=10^3$. Lower panel: The orbit of the center of mass energy. The maximum value reaches $E_{cm}/2m=1.42$ at $t\simeq 400$, where $2m$ is the minimum of $E_{cm}$.
(b) The solution with $\epsilon=10^{-2}, E_0=0.5, R_0=1.9, \sigma_1=1$. The dashed straight line in the lower panel denotes the upper bound, i.e., the maximum when fine tuning is imposed. Twenty-five collisions occur, and the maximum center of mass energy among them is $E_{cm}/2m=4.62$.
\label{fig-sol}}
  \end{center}
\end{figure*}

In \fig{fig-sol} (a), we demonstrate some examples of solutions of \eq{commontime-potential} to see how the shells evolve and the center of mass energy behaves.
We choose $\epsilon=10^{-2}, E_0=0.2, R_0=1.2, \sigma_1=1$ as the initial condition. Note that the parameters are not fine tuned at all.
In this figure, trajectories of the shells are depicted as the red and blue curves in the upper panel, while the orbit of the center of mass energy is shown in the lower panel. 
The time is $t_2$, which is measured between the shells, and we just denote it as $t$ for brevity. 
Since the center of mass energy is normalized by $2m$ (the minimum of $E_{cm}$), its ratio $E_{cm}/2m$ represents the amount of increase in $E_{cm}$ by shell acceleration. 
By defining ${\rm max}E_{cm}$ as the maximum center of mass energy during multiple collisions, we find ${\rm max}E_{cm}/2m=1.42$.
This solution seems to be settled down a stationary motion, and consequently the orbit of the center of mass energy saturates after $t \simeq 400$.

Another solution with initial parameters of $\epsilon=10^{-2}, E_0=0.5, R_0=1.9, \sigma_1=1$ up to $t=10^3$ is plotted in \fig{fig-sol} (b). 
Unlike the solution in \fig{fig-sol} (a), the solution (b) behaves in a complicated way, illustrating no pattern for the orbit of the center of mass energy.

\begin{figure}[htbp]
  \begin{center}
    \begin{tabular}{cc}

      \begin{minipage}{0.5\hsize}
        \begin{center}
          \includegraphics[scale=0.6]{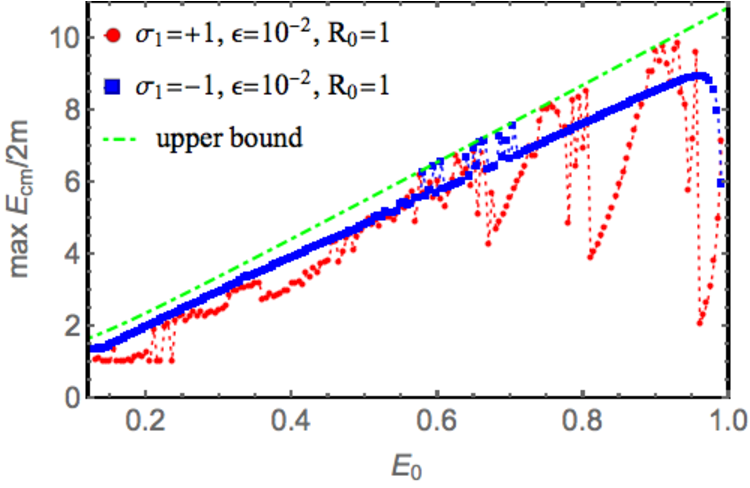}
          (a) 
        \end{center}
      \end{minipage}
      
      \begin{minipage}{0.5\hsize}
        \begin{center}
          \includegraphics[scale=0.6]{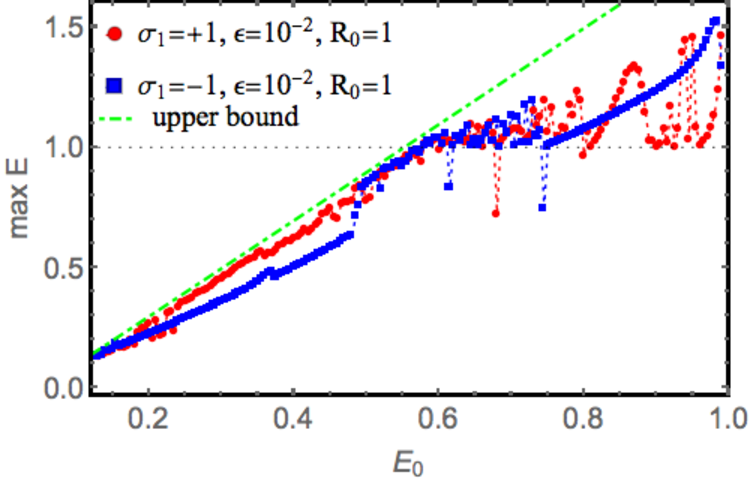}
         (b) 
        \end{center}
      \end{minipage}
      
\vspace{0.4in}  \\
      \begin{minipage}{0.5\hsize}
        \begin{center}
          \includegraphics[scale=0.6]{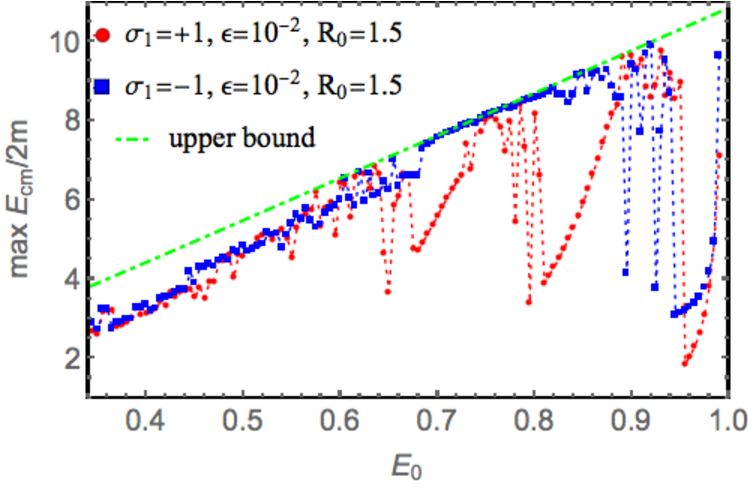}
          (c) 
        \end{center}
      \end{minipage}
     
      \begin{minipage}{0.5\hsize}
        \begin{center}
          \includegraphics[scale=0.6]{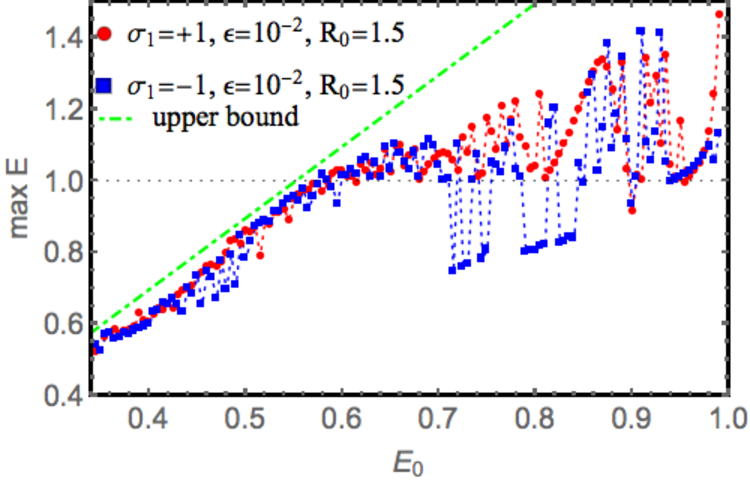}
         (d) 
        \end{center}
      \end{minipage}
 
\\
    \end{tabular}
    \caption{(a) The maximum center of mass energy achieved during multiple collisions as a function of the initial specific energy $E_0$. We take $(\epsilon, R_0)=(10^{-2}, 1)$. The green line denotes the upper bound defined by \eq{thmax-Ecm}. $\sigma_1=\pm1$ is the initial direction of the shell with the radius $R_1$, the initially inner shell.
   (b) The maximum of $E$ as a function of $E_0$ with $(\epsilon, R_0)=(10^{-2}, 1)$. The green line denotes the upper bound of $E$.
   (c) Same as (a) but $R_0=1.5$.
   (d) Same as (b) but $R_0=1.5$.
   }
\label{fig-epsilon2}
  \end{center}
\end{figure}

Now, we solve the maximum center of mass energy as a function of the initial specific energy $E_0$ for $(\epsilon, R_0)=(10^{-2}, 1)$ up to $t=10^4$ \cite{E0range}. This is illustrated in \fig{fig-epsilon2} (a).
We find that the values of the maximum $E_{cm}/2m$ are close to the upper bound defined by \eq{thmax-Ecm}, especially for small $E_0 \lesssim 0.55$, irrespective of the sign of $\sigma_1$. 
For the solution with $\sigma_1=+1$ ($-1$), we find $98.9\%$ ($98.9\%$) out of the upper bound is achieved at $E_0\simeq 0.91$ ($0.60$).
We calculate the arithmetic mean of the ratio of the maximum to the theoretical upper bound, $\langle {\rm max} E_{cm}/E_{cm.b}\rangle$, in order to evaluate how large the observed center of mass energy is achieved compared to the fine tuned center of mass energy. 
As expected from \fig{fig-epsilon2} (a), the mean is considerably high, $\langle {\rm max} E_{cm}/E_{cm.b}\rangle=75.0\%$ for $\sigma_1=+1$ and $\langle {\rm max} E_{cm}/E_{cm.b}\rangle=86.4\%$ for $\sigma_1=-1$. 
This result is important in that a significant proportion of the theoretical maximum is automatically achieved during multiple collisions without fine tuning.
Figure \ref{fig-epsilon2} (b) describes the maximum value of the specific energy $E$ under the same parameter choice as (a). 
The green line is the upper bound for the maximum of $E$ (the explicit form of the function delegated to Appendix \ref{sec:ej-shell-bound}). From this figure we read whether the shell will stay in the system after multiple collisions or eventually one of the shells will escape from the system.
When $E<1$, the shells are confined in the system at least during the integration time. 
To the contrary, when $E\geq1$, one of the two shells gains sufficient energy to escape from the system to infinity. Thus, from \fig{fig-epsilon2} (b), for $E_0 \gtrsim 0.55$, one of the shells in the end escapes to infinity by gaining energy due to collisions. Consequently, the other shell remains and oscillates around the central region. 
We will discuss this mass ejection in detail later.

We also solve equation of motions for the shells with $R_0=1.5$, of which solutions are illustrated in \fig{fig-epsilon2} (c) and \fig{fig-epsilon2}(d). These solutions are qualitatively the same as the case of $R_0=1$, denoting that large center of mass energies are obtained for $E_0\lesssim 0.55$.
For most of the solutions with $E_0\gtrsim 0.55$, one shell lastly gains large energy $E\geq 1$ due to energy exchange by collisions, thereby causing to escape from the system.

\begin{figure}[htbp]
  \begin{center}
    \begin{tabular}{cc}

      \begin{minipage}{0.5\hsize}
        \begin{center}
          \includegraphics[scale=0.6]{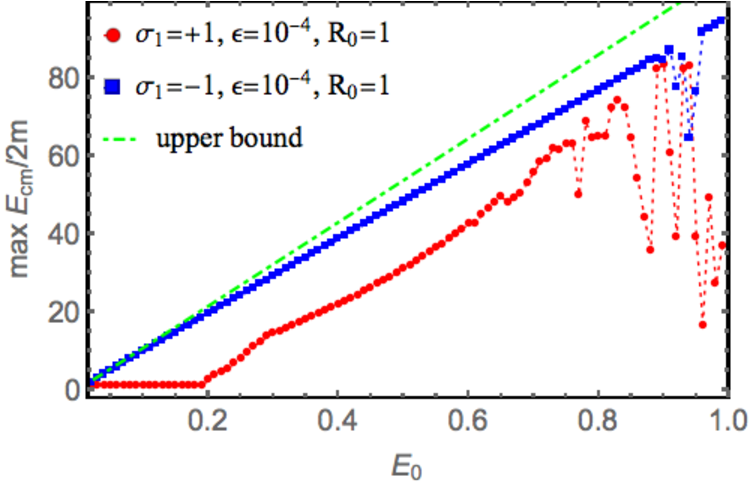}
          (a) 
        \end{center}
      \end{minipage}
      
      \begin{minipage}{0.5\hsize}
        \begin{center}
          \includegraphics[scale=0.6]{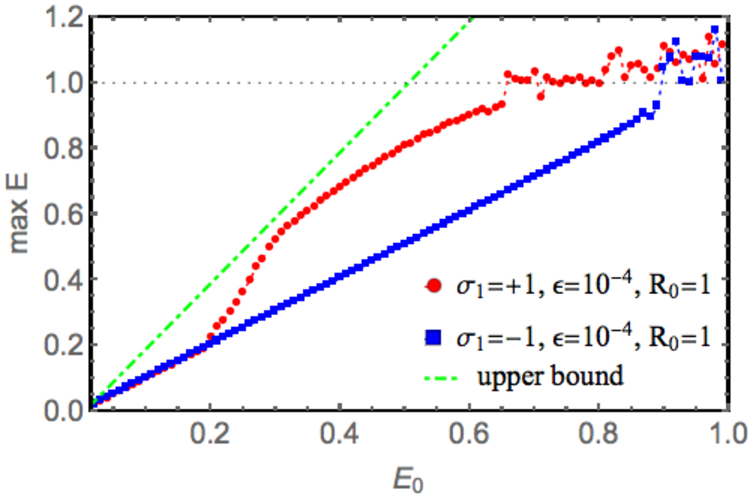}
         (b) 
        \end{center}
      \end{minipage}
      
\vspace{0.4in}  \\
      \begin{minipage}{0.5\hsize}
        \begin{center}
          \includegraphics[scale=0.6]{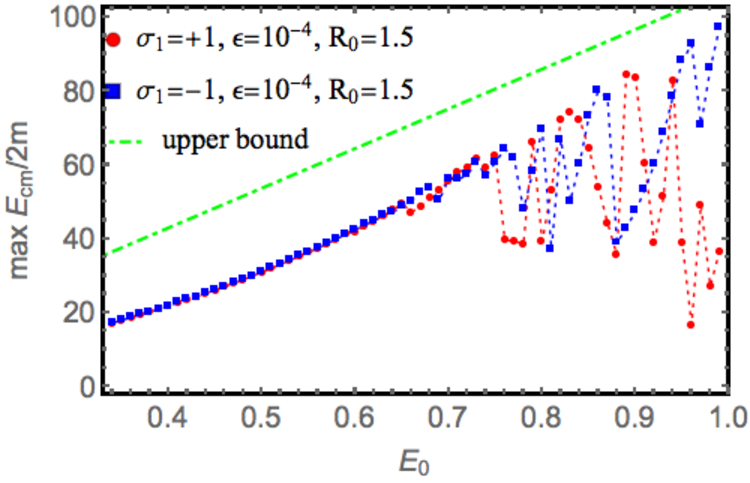}
          (c) 
        \end{center}
      \end{minipage}
     
      \begin{minipage}{0.5\hsize}
        \begin{center}
          \includegraphics[scale=0.6]{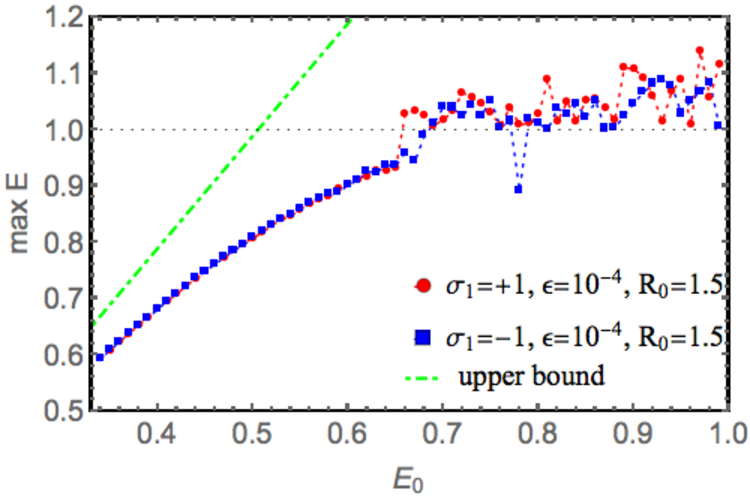}
         (d) 
        \end{center}
      \end{minipage}
 
\\
    \end{tabular}
    \caption{(a) The maximum center of mass energy as a function of $E_0$ with $(\epsilon, R_0)=(10^{-4}, 1)$.  
   (b) The maximum of $E$ as a function of $E_0$ with $(\epsilon, R_0)=(10^{-4}, 1)$.
   (c) Same as (a) but for $R_0=1.5$.
   (d) Same as (b) but for $R_0=1.5$. }
\label{fig-epsilon4}
  \end{center}
\end{figure}

We have taken $\epsilon=10^{-2}$ in the above problem. Since $E_{cm}/2m$ becomes large as $\epsilon$ decreases [see \eq{Ecm-order}], we next solve the problem with the initial condition of $(\epsilon, R_0)=(10^{-4}, 1)$ and $(\epsilon, R_0)=(10^{-4}, 1.5)$ up to $t=5\times10^4$ in \fig{fig-epsilon4}.
These solutions are qualitatively the same as the case of $\epsilon=10^{-2}$. For small energy $E_0$, the two shells exhibit perpetual oscillatory motions because max$E<1$ and large $E_{cm}/2m$ are confirmed, whereas a one shell escapes the system after several collisions for large $E_0$ because of max$E\geq1$.

In Table \ref{tab-table}, we summarize our numerical results on the maximum center of mass energy and the arithmetic mean of the ratio of the maximum to the upper bound. 

\begin{table}[htb]
  \begin{center}
    \caption{The maximum of the center of mass energy and the arithmetic mean of the ratio of the maximum to the theoretical upper bound.}
    \begin{tabular}{p{3em} p{3em} p{5em} p{3em} p{4em} p{3em} p{3em} p{5em} p{3em} } \hline \hline
     \multicolumn{4}{c}{$\epsilon=10^{-2}$} & &  \multicolumn{4}{c}{$\epsilon=10^{-4}$} \\ \cline{1-4} \cline{6-9} 
      $R_0$ & $\sigma_1$ & $\frac{{\rm max}E_{cm}}{2m}$ & $\langle \frac{{\rm max} E_{cm}}{E_{cm.b}}\rangle$  & &    $R_0$ & $\sigma_1$ &  $\frac{{\rm max}E_{cm}}{2m}$ & $\langle \frac{{\rm max} E_{cm}}{E_{cm.b}}\rangle$  \\ \hline 
      1 & 1 & 9.84 & 75.0\%    & &   1 & 1 & 83.5 & 48.9\% \\ \cline{2-4} \cline{7-9}
       & $-1$ & 8.92 & 86.4\% &    &   & $-1$ & 94.4 & 88.7\% \\ \hline
      1.5 & $+1$ & 9.78 & 77.6\% &  &    1.5 & $+1$ & 84.5 & 59.7\% \\ \cline{2-4} \cline{7-9}
       & $-1$ & 9.88 & 84.7\% &     &  & $-1$ & 97.4 & 64.4\% \\ \hline \hline
    \end{tabular}
    \label{tab-table}
  \end{center}
\end{table}

\subsection{Mass ejection}
\begin{figure}[htbp]
  \begin{center}
    \begin{tabular}{cc}
      \begin{minipage}{0.5\hsize}
        \begin{center}
          \includegraphics[scale=0.7]{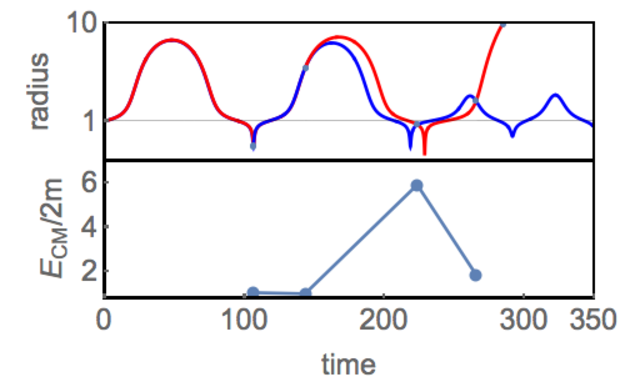}
          (a) 
        \end{center}
      \end{minipage}
      
      \begin{minipage}{0.5\hsize}
        \begin{center}
          \includegraphics[scale=0.7]{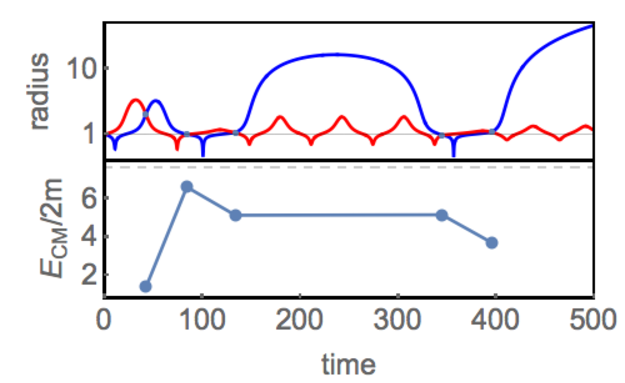}
         (b) 
        \end{center}
      \end{minipage}
 
\\
    \end{tabular}
    \caption{When the initial shell's energy is large enough, mass ejection occurs. (a) The solution with $(\epsilon, R_0, E_0, \sigma_1)=(10^{-2}, 1, 0.85, 1)$, describing shell ejection after four collisions. The shell (the red curve) escapes to infinity, with energy $E=1.24>1$. The remaining shell (the blue curve) continues oscillatory motion.
  (b) The solution with $(\epsilon, R_0, E_0, \sigma_1)=(10^{-2}, 1, 0.7, -1)$. One of the shells is ejected after five collisions. The ejected shell has energy $E=1.13$.}
\label{fig-ejection}
  \end{center}
\end{figure}

Here, we concentrate on the situation where one shell escapes out of the system.
As we have observed, shells are confined for a long time in the system for small $E_0$, whereas one of them can in the end escape to infinity for large $E_0$. We plot some examples of shell ejections for large $E_0$ in \fig{fig-ejection}.

Let us discuss the mass ejection observed in \fig{fig-epsilon2} and \fig{fig-epsilon4}.
See the regions with large initial specific energy $E_0$. All the points lying in the region above the line ${\rm max} E=1$ in Figs. \ref{fig-epsilon2} (b),  \ref{fig-epsilon2} (d), \ref{fig-epsilon4} (b), and \ref{fig-epsilon4} (d) yield that one shell is ejected to infinity after several collisions. 
In \fig{fig-epsilon2} (b) and  \ref{fig-epsilon2} (d), almost all solutions with $E_0 \gtrsim 0.55$ are in the end ejected. In \fig{fig-epsilon4} (b) and  \ref{fig-epsilon4} (d), almost all solutions with larger $E_0$ [$E_0 \gtrsim 0.65$ for $(R_0, \sigma_1)=(1, +1)$ and $(R_0, \sigma_1)=(1.5, \pm1)$, while $E_0 \gtrsim 0.9$ for $(R_0, \sigma_1)=(1,-1)$] are in the end ejected.
This means that shells with initially larger specific energy easily escape from the system.

\begin{figure}[htbp]
  \begin{center}
    \begin{tabular}{cc}
      \begin{minipage}{0.5\hsize}
        \begin{center}
          \includegraphics[scale=0.6]{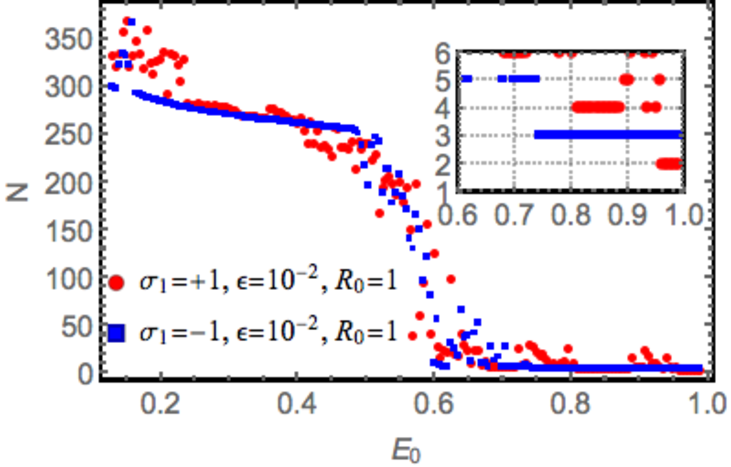}
          (a) 
        \end{center}
      \end{minipage}
      
      \begin{minipage}{0.5\hsize}
        \begin{center}
          \includegraphics[scale=0.6]{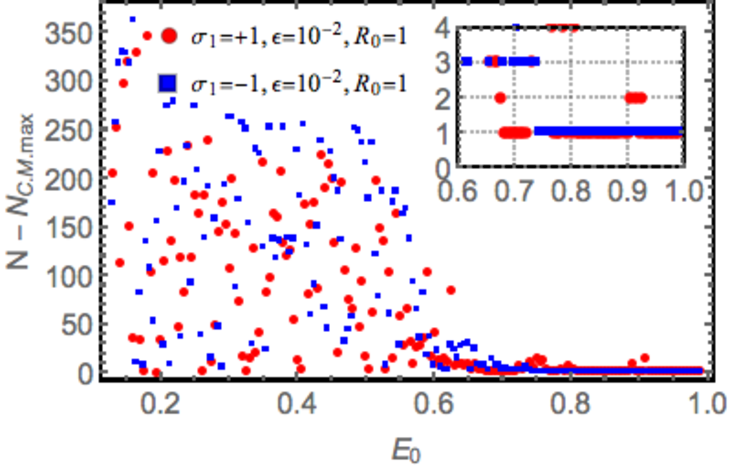}
         (b) 
        \end{center}
      \end{minipage}
\\
    \end{tabular}
    \caption{
    (a) The number of collisions during integration time, as a function of $E_0$ for $(\epsilon, R_0)=(10^{-2}, 1)$. 
    (b) the difference between $N$ and $N_{C.M.max}$ denoting the number of collisions when the maximum $E_{cm}/2m$ is achieved. The difference is $1$ (inset) for almost all solutions with large $E_0$. See the text for the reason.}
\label{fig-epsilon2-N}
  \end{center}
\end{figure}

Now, let us evaluate how many times the shells collide before leaving the system.
Figure \ref{fig-epsilon2-N}(a) denotes the number of collisions during integration time, as a function of $E_0$ for given values of $(\epsilon, R_0)=(10^{-2}, 1)$.
For $E_0\lesssim0.55$, the number of collisions is large because the shells are perpetually  confined in the system. On the contrary, there are few collisions and the number of collisions is at most $\mathcal O(10)$ for $1>E_0\gtrsim0.55$. 
It is known from \fig{fig-epsilon2} (b) that one shell escapes out of the system for the range $1>E_0\gtrsim0.55$, and thus in this range, $N$ means ``the number of collisions before the shell ejection''.

At the same time, we need to see \fig{fig-epsilon2-N} (b), which represents the difference between 
$N$ and $N_{C.M.max}$ denoting the number of collisions when the maximum $E_{cm}/2m$ is achieved.
For $E_0 \lesssim 0.55$, the difference distributes randomly, and no pattern is observed.
On the other hand, for almost all solutions with $E_0\gtrsim0.55$, the difference is unity [see the inset in \fig{fig-epsilon2-N} (b)]. 
This means that the collision with the largest center of mass energy is likely to occur one time before the collision inducing the ejection. This seems to be a universal feature also found in the $\epsilon=10^{-4}$ case.

We want to give a physical meaning on this feature.
For this purpose, we derive the relation between the center of mass energy and the energy transfer between the shells. This can be done simply by eliminating $\gamma$ from \eq{Ecm} and \eq{delta-E}, 
\begin{align}
\Delta E=\frac{m}{R_c}\left(\frac{E_{cm}^2}{2m^2}-1\right).
\label{Ecm-deltaE}
\end{align}
From \eq{Ecm-deltaE}, it is obvious that when the center of mass energy (normalized by $2m$) takes its maximum, the energy transfer also takes its maximum for given $R_c$ and $m$.  
In other words, in a collision with a large $E_{cm}/2m$, large energy transfer also occurs. The maximum energy transfer occurs when the shell collides head on at $R=R_{min}$.
Considering this, when looking at \fig{fig-epsilon2-N} (b), in most cases for large $E_0$, the maximum $E_{cm}/2m$ occurs in the collision one time before the shell escapes. This means that this collision caused a large energy transfer that allows one shell to escape to infinity.
In fact such a case is shown in \fig{fig-ejection}(a).

\subsection{Chaotic nature}
We have observed high energy collision of shells so far. As observed, this high energy phenomenon is evidently caused by random motions of shells. In this section we investigate if the chaotic nature is hidden in our  system. For this purpose we can follow similar procedure given in Ref. \cite{BritoCardosoRocha2016}.

\begin{figure}[htbp]
  \begin{center}
          \includegraphics[scale=0.6]{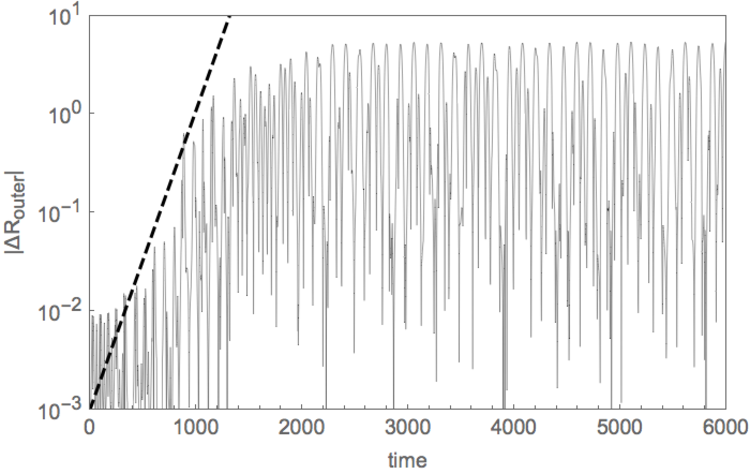}
    \caption{The difference in the orbit of the outermost shell starting from two slightly different radii, $|\Delta R_{{\rm outer}}|$.  We take $(\epsilon, E_0, R_0, \sigma_1)=(10^{-2}, 0.5, 1, 1)$ for the initial parameter of the system. We see an exponential grow of the difference, $|\Delta R_{{\rm outer}}| \propto e^{\lambda t}$ with $\lambda \sim 0.01$ (the dashed line).
\label{fig-deltaRouter}}
  \end{center}
\end{figure}

One of analyses for chaos is the sensitivity on the initial value. Chaos occurs when the distance $x$ between two orbits, starting from slightly different initial values, grows exponentially with time.
This is written as $\Delta x \propto e^{\lambda t}$ with the positive Lyapunov exponent $\lambda$.
If $\lambda$ is negative, this is not a chaos.
In our setup the exponent is determined by following the evolution of the distance between shells starting from slightly different radii. In \fig{fig-deltaRouter}, we plot the difference of the outermost shell's orbits starting from radii that differ initially only by $\Delta R(0)=10^{-3}$. From this figure, an exponential increase in the difference is certainly observed, $|\Delta R_{{\rm outer}}| \propto e^{\lambda t}$ with $\lambda \sim 0.01$. Although the value of $\lambda$ is nearly zero, a positive $\lambda$ is an evidence of chaos. 
The reason for the saturation in the late time seen in \fig{fig-deltaRouter} is that the outermost shell reaches its maximum radius.

\begin{figure}[htbp]
  \begin{center}
          \includegraphics[scale=0.6]{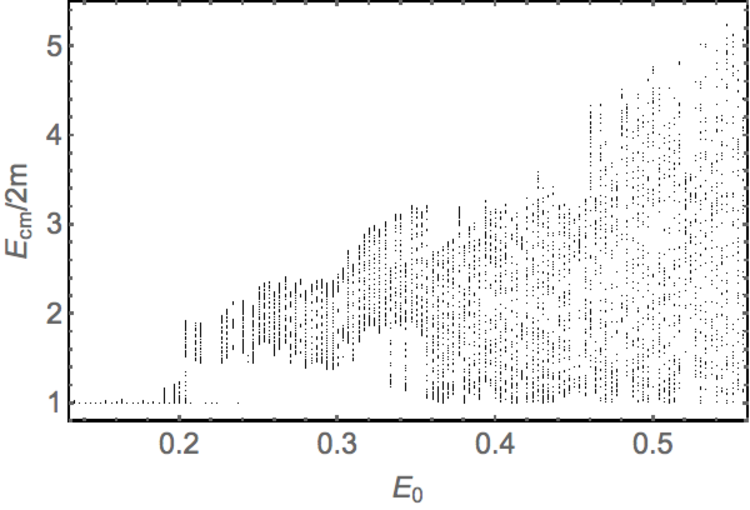}
    \caption{The bifurcation diagram, describing late time orbits of $E_{cm}/2m$ as a function of $E_0$. The orbit nearly converges to unity or is periodic around unity for $E_0 \lesssim 0.2$. As $E_0$ increases, the distribution of the orbits becomes random and continuous. This behavior indicates chaos  in the two-shell system.
\label{fig-bifurcation}}
  \end{center}
\end{figure}

Another method to analyze chaotic behavior is to draw a bifurcation diagram of the center of mass energy. 
The diagram, if the system is chaotic, displays the transition from periodic to nonperiodic orbits of the center of mass energy when we increase a parameter.
For this purpose, we take the initial condition adopted in \fig{fig-epsilon2}(a), $(\epsilon, R_0, \sigma_1)=(10^{-2}, 1, 1)$ and the integration time is set to be $t=10^4$.
Then, the parameter range of $E_0$ for perpetually oscillating solutions is $0.12\lesssim E_0\simeq 0.55$ \cite{E0range}.
In \fig{fig-bifurcation}, the late time orbit (the orbit after $t=7500$) of the center of mass energy is plotted as a function of $E_0$. 
From the figure, we find $E_{cm}/2m$ takes almost the minimum $1$ or oscillates near $1$ for small $E_0$. As $E_0$ increases, the orbit becomes quite complicated and nonperiodic, thereby indicating chaotic nature of the two-shell system.

\section{Conclusion}
\label{sec:conclusion}
We have observed high energy collision of two concentric spherical shells without fine tuning.
We have studied multiple collisions of two dust shells in an overcharged Reissner--Nordstr\"om spacetime, providing repulsive force due to the naked singularity. 
By using the fact that the self-gravity of the shells and repulsive force form a simple gravitationally bound system, we investigated multiple collisions of shells in the bound system when two shells interact only gravitationally. 
We have solved the equation of motions for two shells without imposing any fine tuning of initial parameter on the shells.
Since one shell gives its energy to the other shell every time they collide, time evolutions of the shells with multiple collisions are highly nontrivial. 
Consequently, the center of mass energy for each collision varies nontrivially and even reaches almost its upper bound, the fine-tuned value. 
We confirmed numerically that the maximum value of the center of mass energy reaches (averagely) several tens of percent of the theoretical maximum value (the fine-tuned value) under generic initial conditions.
It was shown that shell ejection occurs when the shell's initial energy is large. Since the center of mass energy and the energy transfer inducing the ejection is simply related through \eq{Ecm-deltaE}, the ejection is likely to occur after the high energy collision is achieved.
The similar ejecting phenomena are found in previous works in Refs. \cite{Barkov+2001MN, Barkov+2002JTEP}.
In these works, energetic ejections occur when the shells collide at a radius as close to the center where the gravitational field becomes stronger. Similarly, in our setup, such ejections can occur when the collision radius is closer to $R_{min}$ where the shell's potential well is the deepest. Thus, our results of shell ejection are in good agreement with previous works. 

We have also found some evidence that behavior of the shells is chaotic.
The chaotic nature found in our system is confirmed in Newtonian \cite{Barkov+2001MN} and in general relativity \cite{Barkov+2002JTEP, BritoCardosoRocha2016}.
Our results indicate that high energy collision occurs in the present bound system thanks to the chaotic nature.

In Newtonian gravity, only the minimum energy transfer takes place at each collision; thereby, no large center of mass energy is achieved [see the text below \eq{delta-E}].
In general relativity, a large amount of energy transfer is possible as they collide at a relativistic speed. That is a unique feature to general relativity.

From the above observations, we conclude that fine tuning for the initial condition of shells is not required for high energy collision in the present bound system.
What we can learn from this study is that, since nature of gravity contains chaos, new physics including high energy phenomena may occur once a bound system inducing the chaotic nature is constructed.


\acknowledgments
The author is grateful to Hongwei Yu, Masashi Kimura, Naoki Tsukamoto, Tomohiro Harada, Tsutomu Kobayashi, Takahisa Igata, Yasutaka Koga, Keisuke Nakashi and Takuya Katagiri for many discussions, suggestions and continuous encouragement.

\appendix

\section{Upper bound of the specific energy of the ejected shell}
\label{sec:ej-shell-bound}
We derive the equation for the upper bound of the specific energy of the ejected shell.
We define the upper bound as the ``total specific energy of the two shells minus the minimum specific energy of the shell that remains in the system''. 
To derive the equation, we first focus on the minimum energy of the other shell, the nonejected shell. 
Since the outer shell is ejected, the nonejected shell is the inner shell. The inner shell has the minimum specific energy $E_c$ of \eq{Ec} by setting $M_+=M_2$ and $M_-=M_1$. 
Since $M_2$ varies between $M_1<M_2<M_3$ by collisions, \eq{Ec} takes its minimum when $M_2$ approaches its maximum $M_3$. Then,
\begin{align}
E_{min}(E_0):=\lim_{M_2\rightarrow M_3} E_c =\sqrt{1-\frac{(M_1+M_3)^2}{4Q^2-m^2}}
=\sqrt{1-\frac{4(1+\mu)^2}{4(1+\epsilon)^2-(\mu/E_0)^2}},
\end{align}
providing that $E_0$ is the initial specific energy of each shell.
In the last equality, we substituted initial parameters in Sec. \ref{sec:initialcondition}.
Next, we focus on total specific energy of the two shells. \eq{energy-cons} guarantees that the sum of the specific energy of two shells is always $2E_0$.
Summarizing above, we derive the upper bound of the specific energy of the ejected shell as a function of $E_0$,
\begin{align}
E_{ej.b}=2E_0-E_{min}(E_0).
\label{Eeject}
\end{align}
The threshold of the ejected energy that the shell can escape to infinity is obtained as $E_{ej.max}=1$. The corresponding threshold $E_0$, satisfying $1=2E_0-E_{min}(E_0)$, is analytically solvable as a quartic equation of $E_0$. However, we can get a simpler and approximated solution by recalling that $0<\epsilon\ll1$ as 
\begin{align}
E_{0}\simeq (1+\sqrt{2-\delta}\sqrt{\epsilon})/2.
\end{align}
From this equation, we find that the minimum specific energy of the ejected shell is a little larger than $0.5$.
For the choice adopted in \fig{fig-epsilon2}, the minimum specific energy is given by $E_0\simeq 0.55$, which sufficiently corresponds to the numerically obtained minimum value as seen in \fig{fig-epsilon2} (b) and \ref{fig-epsilon2} (d).



\begin{thebibliography}{99}
\bibitem{PiranShahamKatz1975}
T. Piran, J. Shaham, and J. Katz, Astrophys. J., \textbf{196}, 107 (1975).

\bibitem{Baushev2008}
A.~N.~Baushev,
Int. J. Mod. Phys. D \textbf{18}, 1195 (2009).

\bibitem{Banados+2010}
M.~Banados, B.~Hassanain, J.~Silk and S.~M.~West,
Phys. Rev. D \textbf{83}, 023004 (2011)

\bibitem{BanadosSilkWest2009}
M.~Banados, J.~Silk and S.~M.~West,
Phys. Rev. Lett. \textbf{103}, 111102 (2009).

\bibitem{Zaslavskii2010}
O.~B.~Zaslavskii,
JETP Lett. \textbf{92}, 571-574 (2010).

\bibitem{Israel1966}
W.~Israel,
Nuovo Cim. B \textbf{44S10}, 1 (1966).

\bibitem{KimuraNakaoTagoshi2010}
M.~Kimura, K.~i.~Nakao and H.~Tagoshi,
Phys. Rev. D \textbf{83}, 044013 (2011).

\bibitem{Patil+2011}
M.~Patil, P.~S.~Joshi, M.~Kimura and K.~i.~Nakao,
Phys. Rev. D \textbf{86}, 084023 (2012).

\bibitem{DrayHooft1985}
T.~Dray and G.~'t Hooft,
Commun. Math. Phys. \textbf{99}, 613-625 (1985).

\bibitem{Nunez+1993}
D.~Nunez, H.~P.~de Oliveira and J.~Salim,
Class. Quant. Grav. \textbf{10}, 1117-1126 (1993).

\bibitem{PoissonIsrael1989}
E.~Poisson and W.~Israel,
Phys. Rev. Lett. \textbf{63}, 1663-1666 (1989).

\bibitem{GasparRacz2010}
M.~E.~Gaspar and I.~Racz,
Class. Quant. Grav. \textbf{27}, 185004 (2010).

\bibitem{NakaoOkawaMaeda2017}
K.~i.~Nakao, H.~Okawa and K.~i.~Maeda,
PTEP \textbf{2018}, no.1, 013E01 (2018).

\bibitem{NakaoYooHarada2018}
K.~i.~Nakao, C.~M.~Yoo and T.~Harada,
Phys. Rev. D \textbf{99}, no.4, 044027 (2019).

\bibitem{Henon1967}
M.~Henon,
Liege International Astrophysical Colloquia \textbf{15}, 243 (1967).

\bibitem{Yangurazova+1984}
L. R. Yangurazova and G. S. Bisnovatyi-Kogan, 
Appl. Sci. Res. \textbf{100}, 319, 17(1984).

\bibitem{MillerYoungkins1997}
B. N. Miller and V. Youngkins, 
Chaos \textbf{7}, 187 (1997).

\bibitem{Barkov+2001MN}
M.~V.~Barkov, V.~A.~Belinski and G.~S.~Bisnovatyi-Kogan,
Mon. Not. Roy. Astron. Soc. \textbf{334}, 338 (2002).

\bibitem{Barkov+2002JTEP}
M.~V.~Barkov, V.~A.~Belinski and G.~S.~Bisnovatyi-Kogan,
J. Exp. Theor. Phys. \textbf{95}, 371-391 (2002).

\bibitem{Barkov+2005}
M. V. Barkov, G. S. Bisnovatyi-Kogan, A. I. Neishtadt, and V. A. Belinski, 
Chaos \textbf{15}, 013104 (2005).

\bibitem{NakaoIdaSugiura1999}
K.~i.~Nakao, D.~Ida and N.~Sugiura,
Prog. Theor. Phys. \textbf{101}, no.1, 47-71 (1999).

\bibitem{IdaNakao1999}
D.~Ida and K.~i.~Nakao,
Prog. Theor. Phys. \textbf{101}, no.5, 989-1000 (1999).

\bibitem{CardosoRocha2016}
V.~Cardoso and J.~V.~Rocha,
Phys. Rev. D \textbf{93}, no.8, 084034 (2016).

\bibitem{BritoCardosoRocha2016}
R.~Brito, V.~Cardoso and J.~V.~Rocha,
Phys. Rev. D \textbf{94}, no.2, 024003 (2016).

\bibitem{MazharimousaviHalilsoy2018}
S. Habib Mazharimousavi and M. Halilsoy,
Int. J. Mod. Phys. D \textbf{27}, 1850064 (2018).

\bibitem{Penrose1969}
R.~Penrose,
Riv. Nuovo Cim. \textbf{1}, 252-276 (1969).

\bibitem{JoshiDwivedi1993}
P.~S.~Joshi and I.~H.~Dwivedi,
Phys. Rev. D \textbf{47}, 5357-5369 (1993).

\bibitem{OriPiran1987}
A.~Ori and T.~Piran,
Phys. Rev. Lett. \textbf{59}, 2137 (1987).

\bibitem{Johnson+1999}
C.~V.~Johnson, A.~W.~Peet and J.~Polchinski,
Phys. Rev. D \textbf{61}, 086001 (2000)

\bibitem{Poisson2009}
E.~Poisson, ``\textit{A Relativist's Toolkit: The Mathematics of Black-Hole Mechanics}'', Cambridge University Press (2007), chap 3.

\bibitem{E0range}
The range of $E_0$ is $\mu/m_0<E_0<1$, where $\mu/m_0$ is defined as the possible minimum energy that the shell can have at the initial time. By substituting $E_c=\mu/m_0$ into \eq{Ec}, we solve
\begin{align}
\mu/m_0=\left\{ \frac{2(Q^2-1-\mu)}{\mu^2}\left( 1-\sqrt{1-\left(\frac{Q\mu}{Q^2-1-\mu}\right)^2} \right)
 \right\}^{-1/2}.
\end{align}

\end{thebibliography}
\end{document}